\documentclass[useAMS,usenatbib]{mn2e}
\usepackage{times}
\usepackage{graphicx}
\title[PSN~J07285387+3349106, a highly reddened Supernova Ibn]{Massive stars exploding in a He-rich circumstellar medium - VIII. PSN~J07285387+3349106, a highly reddened supernova Ibn}
\author[A. Pastorello et al.]{A. Pastorello$^{1}$\thanks{E-mail: andrea.pastorello@oapd.inaf.it}, L. Tartaglia$^{1,2}$, N. Elias-Rosa$^{1}$, A. Morales-Garoffolo$^{3}$,  
\newauthor G. Terreran$^{1,4}$,  S. Taubenberger$^{5,6}$, U. M. Noebauer$^{6}$, S. Benetti$^{1}$, E. Cappellaro$^{1}$,
\newauthor  F. Ciabattari$^{7}$, M. Dennefeld$^{8}$,  A. Dimai$^{9}$, E. E. O. Ishida$^{6}$, A. Harutyunyan$^{10}$,   
\newauthor  S. Leonini$^{11}$,  P. Ochner$^{1}$, J. Sollerman$^{12}$, F. Taddia$^{12}$,  S. Zaggia$^{1}$\\
$^{1}$INAF -- Osservatorio Astronomico di Padova, Vicolo dell'Osservatorio 5, I-35122 Padova, Italy\\
$^{2}$Dipartimento di Fisica e Astronomia, Universit\`a degli Studi di Padova, Vicolo dell'Osservatorio 3, I-35122 Padova, Italy\\
$^{3}$Institut de Ci\`encies de l'Espai (CSIC-IEEC), Campus UAB,  Cam\'i de Can Magrans S/N, 08193 Cerdanyola (Barcelona), Spain\\
$^{4}$Astrophysics Research Centre, School of Mathematics and Physics, Queen's University Belfast, Belfast BT7 1NN, UK\\
$^{5}$European Organisation for Astronomical Research in the Southern Hemisphere (ESO), Karl-Schwarzschild-Str. 2, D-85748 \\Garching bei M\"unchen, Germany\\
$^{6}$Max-Planck-Institut f\"ur Astrophysik, Karl-Schwarzschild-Str. 1, D-85748 Garching bei M\"unchen, Germany\\
$^{7}$Osservatorio Astronomico di Monte Agliale, Via Cune Motrone, Borgo a Mozzano, Lucca, I-55023, Italy\\
$^{8}$Institut d'Astrophysique de Paris, CNRS, and Universite Pierre et Marie Curie, 98 bis Boulevard Arago, 75014, Paris, France\\
$^{9}$Osservatorio Astronomico del Col Drusci\'e,  I-32043 Cortina d'Ampezzo, Italy\\
$^{10}$Fundaci\'on Galileo Galilei-INAF, Telescopio Nazionale Galileo, Rambla Jos Ana Fern\'andez P\'erez 7, E-38712 Bre\~na Baja, TF, Spain\\
$^{11}$Osservatorio Astronomico Provinciale di Montarrenti, S.S. 73 Ponente, I-53018, Sovicille, Siena, Italy\\
$^{12}$The Oskar Klein Centre, Department of Astronomy, Stockholm University, AlbaNova, 10691 Stockholm, Sweden
}
\begin{document}

\date{Accepted YYYY Month DD. Received YYYY Month DD; in original form YYYY Month DD}

\pagerange{\pageref{firstpage}--\pageref{lastpage}} \pubyear{XXXX}

\maketitle

\label{firstpage}

\begin{abstract}

We present spectroscopic and photometric observations for the Type Ibn supernova (SN) dubbed PSN~J07285387+3349106.
Using data provided by amateur astronomers, we monitored the photometric rise of the SN to maximum light, occurred on 
2015 February 18.8 UT (JD$_{max}(V)$ = 2457072.0 $\pm$ 0.8). 
PSN~J07285387+3349106 exploded in the inner region of an infrared luminous galaxy, and is the most reddened SN Ibn discovered so far.
We apply multiple methods to derive the total reddening to the SN, and determine a total colour excess E($B-V$)$_{tot}$ = 0.99 $\pm$ 0.48 mag.
Accounting for the reddening correction, which is affected by a large uncertainty, we estimate a peak absolute magnitude of $M_V$ = $-$20.30 $\pm$ 1.50.
The spectra are dominated by continuum emission at early phases, and He~I lines with narrow P-Cygni profiles are detected.  
We also identify weak Fe~III and N~II features. All these lines show an absorption component which is 
blue-shifted by about 900--1000 km s$^{-1}$. The spectra also show relatively broad He~I line wings with low contrast, 
which extend to above 3000 km s$^{-1}$. From about 2 weeks past maximum,  broad lines
of O~I, Mg~II and the Ca~II near-infrared triplet are identified.
The composition and the expansion velocity of the circumstellar material, and the presence of He~I and $\alpha$-elements in the
SN ejecta indicate that PSN~J07285387+3349106  was produced by the core-collapse of a stripped-envelope star.
We suggest that the precursor was WNE-type Wolf-Rayet star in its dense, He-rich circumstellar cocoon.

\end{abstract}


\section{Introduction}

Type Ibn supernovae (SNe Ibn) are a class of transients whose spectra show narrow lines of He~I 
in emission and weak (or no) evidence of H lines \citep{mat00,pasto08a}. This is interpreted as evidence
for the presence of a He-rich circumstellar medium (CSM). In contrast to SNe IIn, whose CSM is dominated by H,
in SNe Ibn the outer environment is almost H-free. This implies that the progenitor stars of SNe Ibn 
already lost their H envelope, and also (part of) the He-rich layers at the time of their terminal explosion.

As recent publications in the literature have pointed out, the number of SN Ibn discoveries
has significantly grown in the past few years. Modern surveys with
very large fields of view have led to an increase in detection rates and have removed the galaxy selection biases. 
 With deeper imaging, and hence an increased sampled volume, SNe Ibn have also been discovered at relatively
high redshifts. Finally, more accurate typing criteria have significantly decreased the risk of mis-classifications.
In fact, the discoveries of transitional SNe Ibn \citep[including the so-called SNe Ibn/IIn, such as SNe 2005la and 2011hw, 
that both show signatures of some circumstellar H along with He,][]{pasto08b,pasto15a,smith12,mod14,bia14} 
or SNe Ibn that evolve into SNe Ib \citep[e.g. ASASSN-15ed and SN~2015G,][]{pasto15b,fra15}, have shown that 
the classification of an object as a Type Ibn SN is sensitive to the epoch at which the spectrum is obtained.
As a consequence, a proper classification requires good-quality spectra obtained at multiple epochs.

A major problem in SN Ibn identification is that they usually explode in star-forming environments \citep{tad15,pasto15c}, hence their spectra are 
heavily contaminated by the complex host galaxy background.
The rapid post-peak luminosity decline furthermore complicates their accurate characterization.
In fact, most objects faded below the detection limit of optical telescopes within about 2 months \citep{mat00,pasto08b,pasto15d,gor14},
with at least one observed exception \citep[OGLE-2012-SN-006, 
which experienced an unusually slow photometric evolution,][]{pasto15d}. 

The reason for the rapid evolution in the optical bands of most SNe Ibn is debated. 
In several cases, the rapid decline of the optical light curves was linked to a prompt condensation of dust
in SN ejecta or in a cool dense shell \citep[CDS, e.g.][]{chu04,kan12,str12}. In SN~2006jc,  the prototype of this family, this process
was directly observed  \citep{smi08,mat08,dica08,noz08,sak09}. SN~2006jc is a milestone for a number
of reasons. First of all, it is still the only SN Ibn for which a pre-SN outburst was observed (by the amateur astronomer K. Itagaki) 
two years before the SN explosion \citep{nak06,pasto07}. In addition, SN~2006jc was well monitored in
the ultra-violet (UV), optical and near-infrared (NIR) bands since it was
hosted by a nearby galaxy \citep{fol07,pasto07,pasto08a,dica08,anu09,mod14,bia14}.
For this reason, SN~2006jc is still the object that provided the most detailed information on this SN family \citep[][and references therein]{chu09}.

A factor that is expected to affect the detection rate of SNe Ibn, which typically explode  in star-forming galaxies,
is their possible location in a dusty regions. However, no SN Ibn affected by significant reddening has been discovered so far.
The first opportunity to monitor an object of this class with strong line-of-sight extinction is PSN~J07285387+3349106. 
The discovery of this transient by the Lick Observatory Supernova Search (LOSS)\footnote{\it http://w.astro.berkeley.edu/bait/public$\_$html/kait.html}
and its early follow-up information have been reported
on the CBAT {\it ``Transient Object Followup Reports''} pages\footnote{\it http://www.cbat.eps.harvard.edu/unconf/followups/J07285387+3349106.html}.
PSN~J07285387+3349106 was classified by \citet{och15} as a young Type Ibn event based on the presence of narrow emission lines of He~I. 
In addition, the detection of the strong, unresolved  Na~I doublet $\lambda\lambda$5889,5895 (Na~ID) in absorption at the redshift of the host galaxy 
(z = 0.01379) led Ochner et al. to conclude that the SN was significantly reddened, with E($B-V$)$_{tot}$ $\sim$ 1 mag.
Multi-band optical photometry of PSN~J07285387+3349106 has been published by \citet{tsv15}.

This article is the eighth of a series whose main goals are to illustrate the variety of physical properties that characterize the Type Ibn SNe family,
and to constrain the nature of their progenitor stars. This paper is arranged as follows: In Section \ref{host}, we describe the main properties
of the galaxy hosting PSN~J07285387+3349106. In Section \ref{reddening}, we compute the total reddening in the SN direction. In Section \ref{redu},
we present the observations and illustrate the data reduction procedures. Sections \ref{lc}~and \ref{spec}  analyse the SN light curve and the 
spectroscopic evolution, respectively. A discussion and a summary follow in Section \ref{disc}.

\begin{figure}
\includegraphics[width=9cm,angle=0]{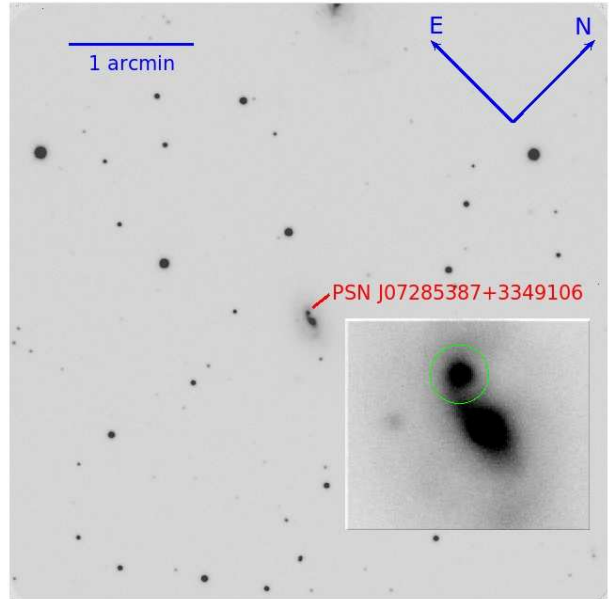} 
 \caption{PSN~J07285387+3349106 and its host galaxy. $R$-band image taken on 2015 October 2 with ALFOSC, mounted at the Nordic Optical Telescope. 
The insert shows a blow-up of the SN region. \label{fig1}}
\end{figure}

\section{The host galaxy} \label{host}

The galaxy hosting PSN~J07285387+3349106, NGC 2388 (Figure \ref{fig1}), is classified as an Sb-type in Hyperleda\footnote{\it http://leda.univ-lyon1.fr/}
\citep{mak14},
and is apparently one member of the galaxy group WBL 142, which is composed of three galaxies within a radius of 82 arcmin \citep{zwi61-68}, 
 whose mean redshift is z $\sim$ 0.0135 \citep{whi99}. 
NGC 2388 has been classified as a possible luminous infrared galaxy (LIRG)
by \citet{san03}. As it has been detected from the UV to the radio wavelengths, its spectral energy distribution is well
constrained \citep{bro14}. In addition, our SN follow-up spectra (see Section \ref{spec}) show very prominent emission lines attributed to
the host galaxy. All of this suggests that NGC 2388 is a star-forming galaxy.

The radial velocity of NGC 2388 corrected for Local Group infall onto Virgo is v$_{Vir}$ = 4255 km s$^{-1}$ \citep[from Hyperleda,][]{mak14}. Adopting a standard cosmology with 
H$_0$ = 73 km s$^{-1}$ Mpc$^{-1}$, $\Omega_M$ = 0.27, $\Omega_\Lambda$ = 0.73,
we obtain a luminosity distance of 58.9 $\pm$ 4.1 Mpc \citep[see][for information on the cosmology calculator]{wri06},
hence a distance modulus $\mu$ = 33.85 $\pm$ 0.15 mag.\footnote{The errors on the distance and distance modulus were computed accounting for the uncertainties on the recessional velocity, 
the Local Group infall correction, and the cosmological parameters, following the prescriptions of the NED database, {\it https://ned.ipac.caltech.edu/}.}

As the total apparent magnitude of NGC 2388 provided by Hyperleda is $B$ = 14.66 $\pm$ 0.32, the total absolute  $B$-band magnitude corrected for Galactic and
internal extinction (following the prescriptions of Hyperleda) is $-$19.76. 
Using the luminosity vs. metallicity relations provided by \citet{tre04}, we 
estimate an integrated oxygen abundance of 12 + log(O/H) = 8.91 (dex) for NGC 2388.
The SN exploded in the proximity of a spiral arm of the host galaxy, about 6 arcsec East and 2 arcsec North of the nucleus. 
Accounting for the SN position, and following \citet{pasto15c}, we estimate the $R_{0,SN}/R_{25}$ ratio\footnote{R$_{SN}$ is 
the deprojected position of the SN, while R$_{25}$ is the isophotal radius for the $B$-band surface brightness of 25 mag arcsec$^{−2}$.}  
to be 0.26. This allows us to compute the oxygen abundance at the SN position, assuming a radial dependence as in \citet{pil04}.
We obtain 12 + log(O/H)$_{SN}$ = 8.81 (dex), which is slightly above the average oxygen abundance estimated
at the SN location for the host galaxies of SNe Ibn by  \citet[][8.63 $\pm$ 0.42 dex]{pasto15c} and  \citet[][8.45 $\pm$ 0.10 dex]{tad15}.

The analysis of the host galaxy of PSN~J07285387+3349106 confirms the results of the above systematic studies, namely that
there is no evidence of SNe Ibn preferentially exploding in metal-poor environments.

\section{The line-of-sight extinction to PSN~J07285387+3349106} \label{reddening}

\begin{table*}
 \centering
 \begin{minipage}{165mm}
  \caption{Photometric data and associated errors for PSN~J07285387+3349106. \label{photo_tab}}
  \begin{tabular}{cccccccccc}
  \hline
Date        &  JD       &  $B $         &   $V$         &   $u$     &  $g$      & $r$          &  $i$      &  $z$      & Instrument  \\
 \hline
20141212  & 2457003.53   &      --       &  $>$19.15     &    --      &    --      & $>$19.03     &    --       &    --      &   $1$ \\
20141223  & 2457014.42   &      --       &  $>$19.01     &    --      &    --      & $>$18.80     &     --      &     --     &   $2$ \\
20141226  & 2457018.46   &      --       &  $>$19.88     &    --      &    --      & $>$19.76     &    --       &    --      &   $1$ \\
20141227  & 2457018.63   &      --       &  $>$20.41     &    --      &    --      & $>$20.19     &    --       &    --      &   $3$  \\
20150107  & 2457030.49   &      --       &  $>$19.10     &    --      &    --      & $>$18.90     &     --      &     --     &   $2$ \\
20150112  & 2457035.49   &      --       &  $>$20.19     &    --      &    --      & $>$19.97     &    --       &    --      &   $3$  \\
20150209  & 2457063.47   &      --       &  18.84  (0.35)  &    --      &    --      & 18.62  (0.26)  &    --       &    --      &   $3$  \\
20150210  & 2457064.45   &      --       &  18.29  (0.38)  &    --      &    --      & 18.14  (0.30)  &    --       &    --      &   $4$ \\ 
20150214  & 2457068.37   &      --       &  17.09  (0.29)  &    --      &    --      & 16.73  (0.32)  &    --       &    --      &   $5$ \\  
20150217  & 2457071.33   &      --       &     --        &    --      &    --      & 16.41  (0.24)  &    --       &    --      &   AFOSC \\
20150217  & 2457071.34   &      --       &  16.65  (0.24)  &    --      &    --      & 16.46  (0.26)  &    --       &    --      &   $1$ \\
20150218  & 2457072.26   &   17.46 (0.09)  &  16.62  (0.09)  &    --      &    --      &    --        &    --       &    --      &   AFOSC \\
20150219  & 2457072.39   &       --      &      --       & 18.69 (0.05) & 17.28 (0.03) & 16.48 (0.03)   &  16.20 (0.04) & 16.01 (0.06) &   AFOSC  \\ 
20150220  & 2457073.52   &   17.77 (0.14)  &  16.71  (0.09)  & 18.85 (0.08) & 17.30 (0.06) & 16.53 (0.06)   &  16.24 (0.04) &    --      &   AFOSC  \\
20150222  & 2457075.58   &       --      &       --      &    --      & 17.45 (0.09) & 16.58 (0.09)   &  16.39 (0.08) & 16.00 (0.16) &   ROSS2  \\
20150222  & 2457076.46   &   17.99 (0.03)  &  16.92  (0.02)  &    --      &      --    & 16.71 (0.03)   &  16.47 (0.05) &    --      &   ALFOSC \\
20150223  & 2457077.48   &   18.09 (0.07)  &  17.11  (0.06)  & 19.46 (0.08) & 17.69 (0.06) & 16.79 (0.08)   &  16.49 (0.06) &    --      &   LRS \\
20150225  & 2457078.54   &       --      &       --      &    --      & 17.83 (0.10) & 16.85 (0.14)   &  16.58 (0.09) & 16.17 (0.13) &   ROSS2 \\ 
20150225  & 2457078.58   &   18.20 (0.07)  &  17.17  (0.09)  &    --      &      --    &      --      &      --     &     --     &   LRS \\
20150227  & 2457081.36   &   18.70 (0.20)  &  17.64  (0.21)  &    --      &      --    &      --      &      --     &     --     &   MEIA \\
20150228  & 2457082.39   &       --      &       --      &    --      &      --    & 17.30  (0.25)  &      --     &     --     &   LRS \\
20150301  & 2457082.54   &       --      &       --      &    --      & 18.09 (0.26) & 17.30  (0.15)  &  16.99 (0.15) & 16.39 (0.21) &   ROSS2  \\ 
20150302  & 2457083.59   &       --      &       --      &    --      & 18.25 (0.23) & 17.54  (0.20)  &  17.11 (0.16) & 16.47 (0.26) &   ROSS2 \\ 
20150302  & 2457084.37   &   19.12 (0.38)  &  18.13  (0.23)  &    --      &     --     &      --      &      --     &     --     &   MEIA \\
20150303  & 2457085.47   &      --       &       --      &    --      &     --     & 17.80  (0.22)  &      --     &     --     &   ALFOSC \\
20150304  & 2457085.67   &      --       &       --      &    --      & $>$16.99   & $>$16.42     &  17.43 (0.53) & $>$15.55   &   ROSS2  \\
20150304  & 2457086.32   &   $>$17.99    &       --      &    --      &     --     &      --      &      --     &     --     &   MEIA \\
20150305  & 2457086.53   &      --       &      --       &    --      & 18.96 (0.36) & 18.02 (0.23)   &  17.59 (0.19) & 17.13 (0.29) &   ROSS2  \\
20150307  & 2457089.35   &      --       &  19.21  (0.42)  &    --      &    --      & 18.96 (0.39)   &      --     &     --     &   $3$  \\
20150308  & 2457090.35   &      --       &  19.59  (0.42)  &    --      &    --      & 19.27 (0.35)   &      --     &     --     &   $3$ \\
20150310  & 2457091.54   &      --       &      --       &    --      & 20.35 (0.51) & 20.00 (0.59)   &  19.34 (0.46) & 18.97 (0.38) &   ROSS2  \\
20150310  & 2457092.30   &      --       &  $>$19.04     &     --     &     --     & $>$18.83     &     --      &     --     &   $2$ \\
20150310  & 2457092.47   &   21.89 (0.23)  &  20.42  (0.22)  &     --     &     --     & 20.13 (0.30)   &  19.69 (0.34) &     --     &   ALFOSC \\
20150311  & 2457093.30   &      --       &      --       & 22.51 (0.54) & 21.08 (0.20) & 20.49 (0.19)   &  19.89 (0.14) & 19.39 (0.12) &   AFOSC  \\  
20150312  & 2457094.33   &      --       &  $>$20.23     &     --     & 21.35 (0.19) & 20.60 (0.18)   &  20.04 (0.14) & 19.58 (0.15) &   AFOSC \\
20150330  & 2457112.34   &      --       &  $>$20.02     &     --     &     --     & $>$19.78     &     --      &     --     &   $3$ \\
20150330  & 2457112.34   &      --       &  $>$18.01     &     --     &     --     & $>$17.85     &     --      &     --     &   $1$ \\
 \hline
\end{tabular}

$1 -$ 0.50-m Newton telescope (Lotti) equipped with a FLI Proline 4710 camera with E2V CCD47-10 (Osservatorio Astronomico di Monte Agliale, Borgo a Mozzano, Lucca, Italy); 
$2 -$ 0.28-m Schmidt-Cassegrain telescope (Vittore Maioni) with a SBIG ST-8 3 CCD Camera (Osservatorio Astronomico del Col Drusci\'e, Cortina, Belluno, Italy); 
$3 -$ 0.53-m Ritchey-Chr\'etien telescope equipped with an Apogee Alta U47 camera with E2V CCD47-10 (Osservatorio Astronomico Provinciale di Montarrenti, Siena, Italy); 
$4 -$ 0.40-m Newton Marcon telescope equipped with an ATIK 428EX camera with Sony ICX674 CCD (private observatory; obs. Paolo Campaner, Ponte di Piave, Treviso, Italy); 
$5 -$ 0.14-m TEC refractor telescope equipped with an ATIK 460EX camera with Sony ICX694ALG CCD and Luminance filter (private observatory; obs. Manfred Mrotzek, Buxtehude, Germany).
\end{minipage}
\end{table*}

\begin{table*}
 \begin{minipage}{165mm}
  \caption{Log of spectroscopic observations of PSN~J07285387+3349106. \label{spec_tab}}
  \begin{tabular}{cccccccccc}
  \hline
Date        &  JD &  Phase (d)$^\ddag$ &   Instrumental configuration    &  Exposure time (s) &  Range (\AA) & Resolution  (\AA) \\
 \hline
2015/02/17 & 2457071.35 & -0.6 & 1.82-m Copernico + AFOSC + gm4 & 1800 & 3500--8190 & 13 \\
2015/02/18 & 2457072.25 & +0.3 & 1.82-m Copernico + AFOSC + gm4 + VPH6 & 1800 + 1800 & 3550--9270 & 14+15 \\
2015/02/19 & 2457073.49 & +1.5 & 1.82-m Copernico + AFOSC + gm4 &  2 $\times$ 1800 & 3440--8170 & 13 \\
2015/02/22 & 2457076.48 & +4.5 & 2.56-m NOT + ALFOSC + gm4 & 1800 & 3330--9100 & 18 \\
2015/02/23 & 2457077.43 & +5.4 & 3.58-m TNG + LRS + LRB + LRR & 1500 + 1500 & 3450-10190 & 10+10 \\
2015/02/25 & 2457078.60 & +6.6 & 3.58-m TNG + LRS + LRB & 1029 & 3450--8060 & 10 \\
2015/02/28 & 2457082.45 & +10.5 & 3.58-m TNG + LRS + LRB & 1800 & 3450--8060 & 10 \\
2015/03/03 & 2457085.49 & +13.5 & 2.56-m NOT + ALFOSC + gm4 & 2700 & 3500--9110 & 18 \\
2015/03/11 & 2457092.51 & +20.5 & 2.56-m NOT + ALFOSC + gm4 &  2 $\times$ 2700 & 3300--9110 & 18 \\ 
 \hline
\end{tabular}
$^\ddag$  The phases are with respect to the $V$-band maximum light (JD$_{max}(V)$ = 2457072.0).
\end{minipage}
\end{table*}

As stated by \citet{och15}, PSN~J07285387+3349106 is significantly extinguished, and hence the estimate of its reddening is of utmost importance.
The Milky Way contribution to the total reddening is modest. Following the revised Galactic reddening estimates of \citet{sch11} and
assuming a \citet{car89} reddening law with R$_V$ = 3.1, we obtain E($B-V$)$_{MW}$ = 0.05 mag.

Obtaining the SN extinction in the host galaxy is more complicated. For this purpose, we measure 
the equivalent width (EW) of the narrow interstellar Na~I $\lambda\lambda$5890,5896 (Na~ID) in the SN spectra to improve the preliminary estimate reported in \citet{och15}.
Unfortunately, this measurement is uncertain, because the region is affected by the the presence of a SN CSM emission line (He~I $\lambda$5876, see Section \ref{spec}) which
increases in strength with time. 
Since  the He~I line is still very weak in our earliest spectrum, it can be used to derive a more robust 
estimate despite its relatively poor signal-to-noise (S/N). From this, we measure
an  EW = 6.5 $\pm$ 0.5 \AA~for the interstellar Na~ID. According to the empirical relation 
provided by \cite{tur03}, we obtain E($B-V$)$_{host}$ = 1.03 mag. 
However, it is known that the relation between Na~ID EW and E($B-V$) has a large dispersion, especially when the interstellar extinction is high \citep[see, e.g.,][]{poz11}. 
In order to account for this, we note that for EW larger than 4~\AA, the SN sample of Poznanski et al. spans a range of $A_V$ values from about 0.5 to 3 mag.
Therefore $\Delta A_V \approx 1.25$ max can be considered an indicative guess for the error on the reddening with this method.
Hence, E($B-V$)$_{tot}$ = 1.08 $\pm$ 0.40 mag is our estimate for the total interstellar extinction to PSN~J07285387+3349106 using the interstellar Na~ID.

As the above method is affected by a large uncertainty, we estimate the amount of extinction
using alternative methods. First, we follow \citet{cal94} and compute the total reddening to the SN through the host galaxy Balmer lines decrement. 
The decrement is measured using different spectra extracted in regions very close to the SN position. With this approach, we infer values 
of E($B-V$)$_{tot}$ $\approx$ 1 to 1.5 mag (with a weighted average of 1.34 $\pm$ 0.31 mag). 

Another method to compute the foreground reddening, which is widely used for Type Ia SNe \citep[e.g.,][]{phi99}, is through the comparison of SN colour curves
at defined epochs. In the case of Type Ibn SNe, this method is limited by the large heterogeneity of this class of transients. However, using the intrinsic
colours of a limited number of SNe Ibn which are photometrically similar to PSN~J07285387+3349106 (see Section \ref{lc} for more details), we obtain a somewhat 
smaller value for the reddening, viz. E($B-V$)$_{tot}$ = 0.81 $\pm$ 0.21 mag.

A weighted average of these three methods provides as best estimate of the total reddening E($B-V$)$_{tot}$ = 0.99 $\pm$ 0.48 mag, which will be used throughout the paper.

\section{Observations and data reduction} \label{redu}

Our spectroscopic and photometric follow-up campaigns were triggered soon after the classification. Early-time photometry data and pre-discovery observations of the field were
gathered a posteriori thanks to the routine unfiltered observations of NGC 2388 by a number of amateur astronomers in the framework of the Italian Supernovae Search Program 
(ISSP)\footnote{{\it http://italiansupernovae.org}; the ISSP is a coordinated SN search involving several Italian Amateur Observatories located in Veneto, Toscana and 
Lombardia. Since 2011 it has led to the discovery of about 80 SNe.}. Details on the observations obtained using 30- to 50-cm telescopes available to the ISSP collaboration 
are provided in the footnotes of Table \ref{photo_tab}. 

Standard photometric follow-up of PSN~J07285387+3349106 was performed using the following instruments: the 1.82-m Copernico Telescope of the INAF - Padova Observatory at Mt. Ekar 
(Asiago, Italy) equipped with AFOSC; the 2.56-m Nordic Optical Telescope (NOT) with ALFOSC and  the 3.58-m Telescopio Nazionale Galileo (TNG) equipped with DOLORES (LRS), both
located at Roque de los Muchachos Observatory (La Palma, Canary Islands, Spain); the 0.8-m Joan Or\'o Telescope (TJO) of the Observatori Astron\`omic del Monsec (Catalonia, Spain) equipped with MEIA;
the INAF 0.6-m Rapid Eye Mount (REM) Ritchey-Chr\'etien reflector Telescope hosted by the European Southern Observatory (ESO, La Silla, Chile) equipped with ROSS2.

\begin{figure*}
\includegraphics[width=11.5cm,angle=0]{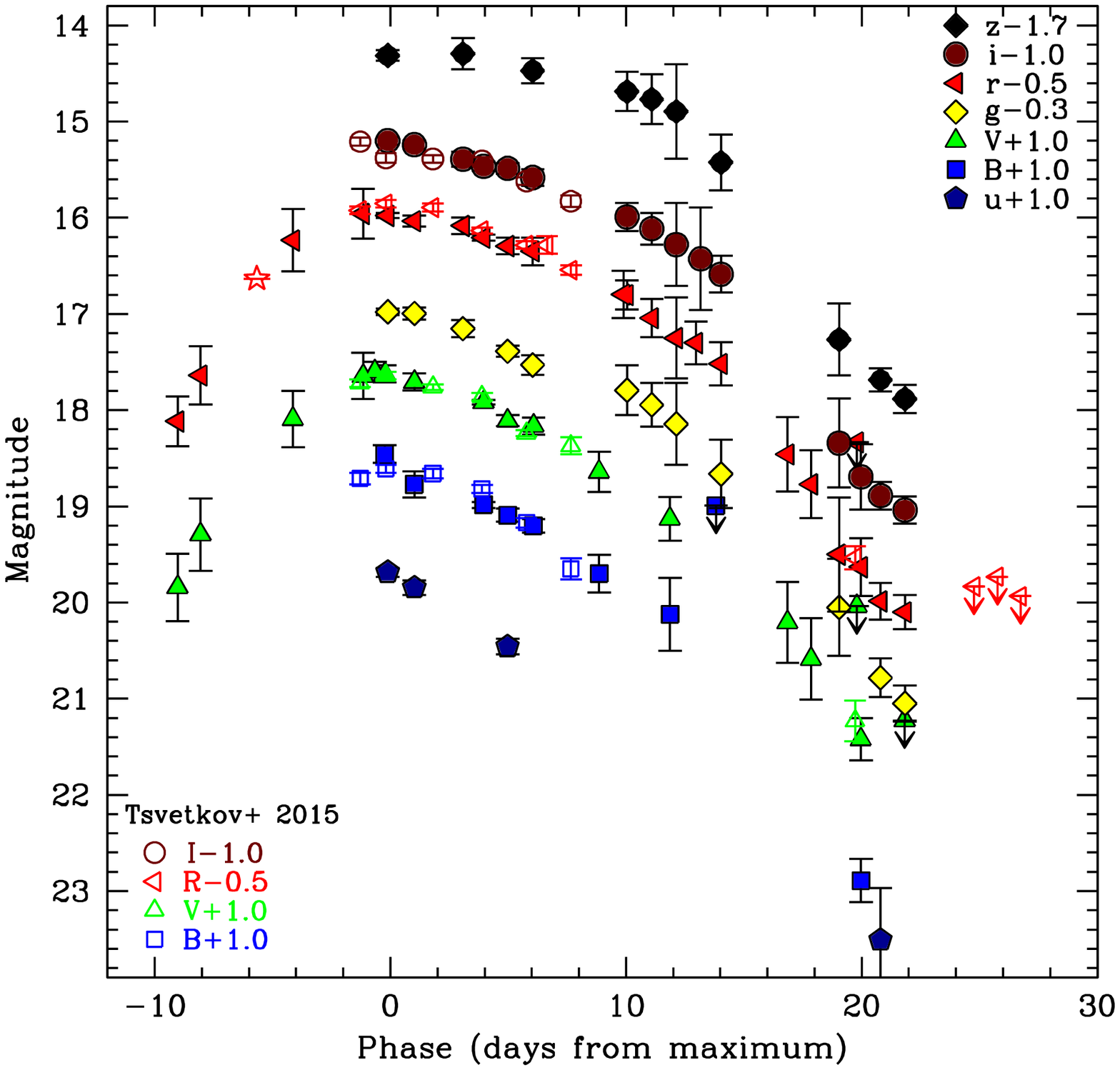} 
 \caption{Multi-band light curves of PSN~J07285387+3349106: our observations (filled symbols) are compared with those published by \protect\citet[][open symbols]{tsv15}.
A constant zeropoint correction has been applied to the Johnson-Cousins $R$- and $I$-band data of  \protect\citet{tsv15} to convert the
magnitudes into the Sloan $r$- and $i$ system. The corrections terms, $\Delta R = 0.23 \pm 0.03$ mag and $\Delta I = 0.47 \pm 0.03$ mag,
have been obtained by computing the synthetic photometry from the available spectra of PSN~J07285387+3349106. The starred symbol represents the
scaled discovery photometry from LOSS. \label{fig2}}
\end{figure*}

Photometric images were pre-reduced using standard tasks available in \textsc{IRAF}\footnote{IRAF is distributed by the National Optical Astronomy Observatory,
which is operated by the Association of Universities for Research in Astronomy (AURA) under cooperative agreement with the National
Science Foundation.}. The pre-reduction steps include overscan, bias and flat-field corrections, and final trimming of the useless image regions.
Using the dedicated  \textsc{PYTHON} pipeline \textsc{SNOoPY} \citep{ec14},\footnote{\textsc{SNOoPY} 
is a collection of \textsc{PYRAF} programs and other widely used public tools (e.g. \textsc{DAOPHOT}, \textsc{SEXTRACTOR}, \textsc{HOTPANTS}).} we then astrometrically registered
the images, extracted a number of stellar sources in the images and measured their instrumental magnitudes using a PSF-fitting technique. As the SN was located in a complex
region inside the host galaxy, a properly removal of the background contribution using templates available in public archives (e.g. SDSS for the Sloan-band images) was required.
The SN magnitude was measured after the host galaxy contamination was removed.

The magnitudes of a number of stars in the field of PSN~J07285387+3349106 are available in the SDSS catalogue. These were used to estimate zeropoints and colour terms 
for all nights for which Sloan-band observations (in $u$, $g$, $r$, $i$, and $z$) were obtained. The Johnson $B$ and $V$ magnitudes were instead calibrated by converting 
the Sloan magnitudes of the stars in the SN field to Johnson magnitudes, following \citet{cho08}.

Early unfiltered data, kindly provided by amateur astronomers, were converted to Johnson $V$ and Sloan $r$ magnitudes. 
The calibrated SN magnitudes are listed in Table \ref{photo_tab}. The reported errors account for 
the uncertainties in the instrumental magnitudes and the photometric calibration. Information on the instruments used by amateur astronomers
is provided in the footnotes of Table~\ref{photo_tab}.

\begin{figure*}
\includegraphics[width=16cm,angle=0]{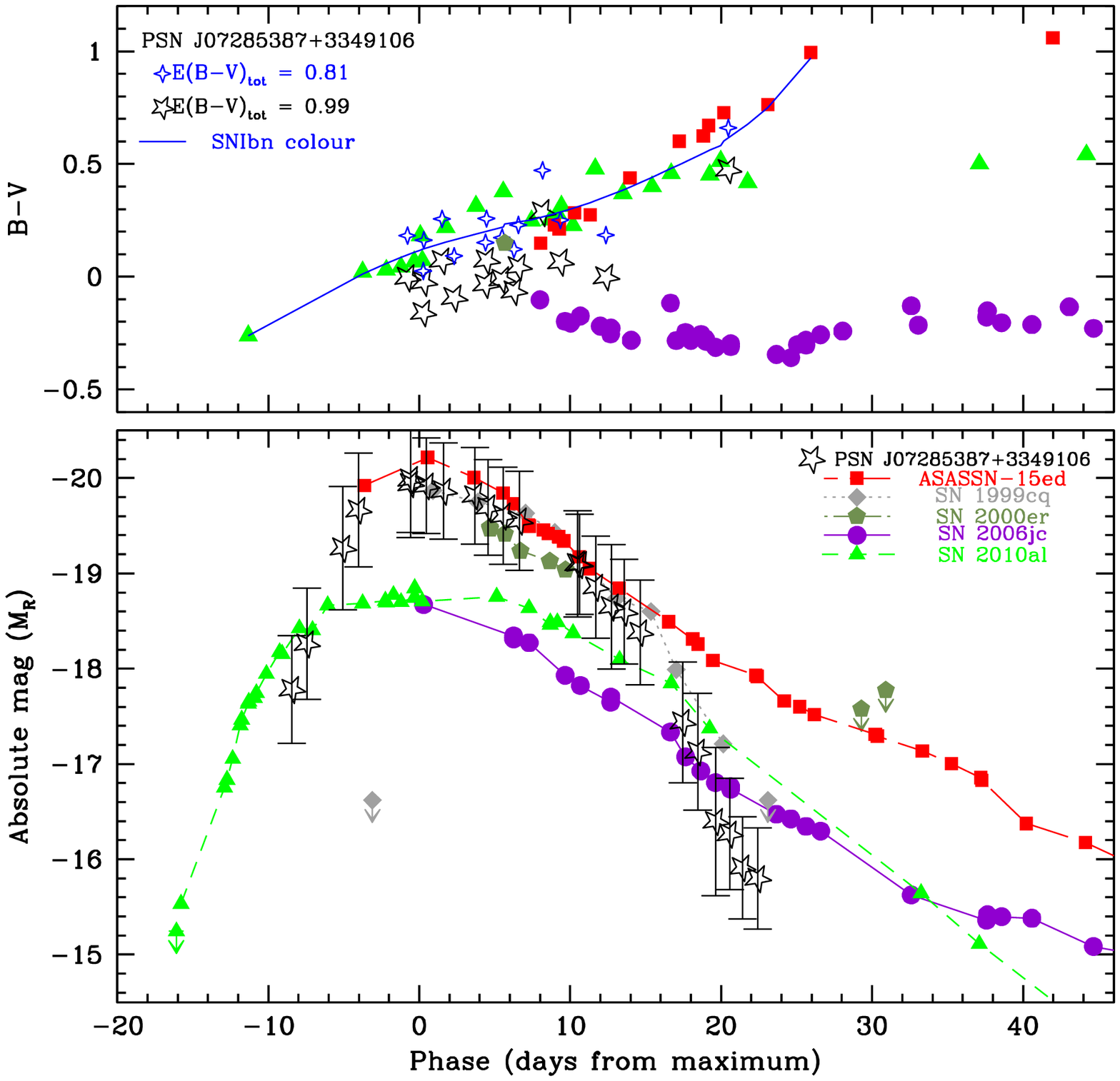} 
 \caption{Top -- Comparison of the $B-V$ colour evolution  of PSN~J07285387+3349106 with those of a sub-sample of SNe Ibn,
including the photometrically similar SNe 2000er \citep{pasto08a} and ASASSN-15ed \citep{pasto15b}, and the well-followed SNe 2006jc \citep{pasto07,fol07} and 2010al \citep{pasto15a}. 
 The  blue solid line represents our SN Ibn colour template (see text).
We show the colour evolution of  PSN~J07285387+3349106 for two values of E($B-V$)$_{tot}$: 0.99 mag, as adopted throughout the paper, 
and  0.81 mag, obtained assuming that PSN~J07285387+3349106 has the same colour evolution as the SN Ibn colour template.
Bottom -- Comparison of the $R$-band absolute light curve of PSN~J07285387+3349106 (obtained from the Sloan-r light curve scaled by -0.23 mag and corrected for E($B-V$)$_{tot}$ = 0.99 mag), with those of the same sample as 
above plus SN~1999cq \citep{mat00}. 
The LOSS discovery magnitude of PSN~J07285387+3349106 is also included. 
The error bars account for the uncertainties of the photometric data, the interstellar extinction and the distance modulus. 
\label{figabs}}
\end{figure*}

Spectroscopic data, covering a 3-week temporal window starting at maximum light, were obtained with some of the instruments mentioned above, including AFOSC, 
mounted on the Copernico Telescope, LRS on the TNG and ALFOSC on the NOT.  
These data were processed using standard IRAF tools. Two-dimensional spectra were bias, flat-field and overscan corrected, afterwards one-dimensional spectra were
extracted from the images. The spectra were then wavelength calibrated using arc spectra obtained with the same instrumental configuration
as the SN spectra.  The wavelength calibration was accurately verified through a cross-check with the position of selected night sky lines. When necessary, 
we applied a constant wavelength shift.
The flux calibration was performed using spectra of flux standard stars observed in the same night as the SN, or using instrumental sensitivity functions
for the same instrumental set-up available in our archive. Then, using the available broad-band photometry, the  
flux calibration of the SN spectra was verified and, in case of a discrepancy with the photometry, the spectra were multiplied by a constant factor. 
Contamination from O$_2$ and OH telluric bands was removed  using the spectra of standard stars.
Information on the spectra of PSN~J07285387+3349106 and the instrumental configurations is listed in Table \ref{spec_tab}.

\section{Light curve} \label{lc}

\begin{figure*}
\includegraphics[width=12cm,angle=270]{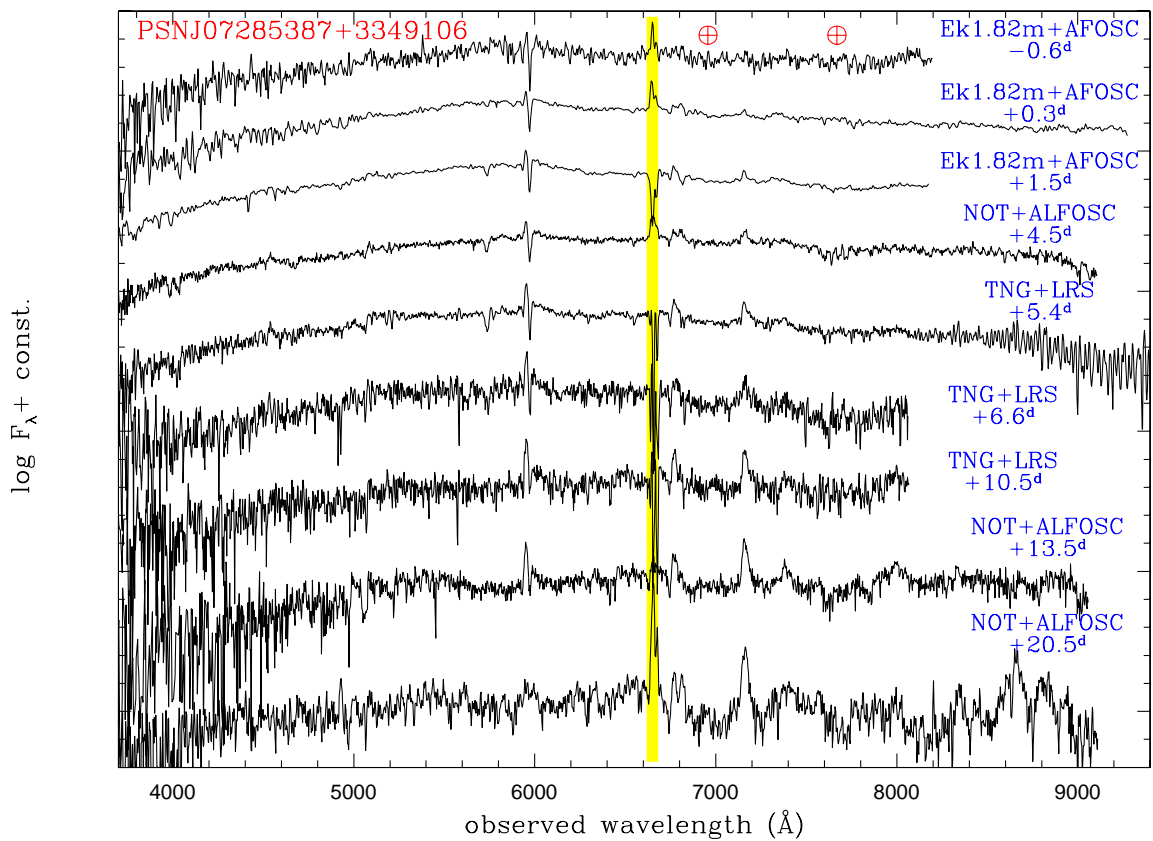} 
 \caption{Spectral sequence of PSN~J07285387+3349106. The region of H$\alpha$ is largely contaminated by the narrow H$\alpha$ and  [N~II] lines from the host galaxy background (marked by the yellow vertical band). No reddening or redshift corrections have been applied to the spectra. \label{figspec}}
\end{figure*}

In Figure \ref{fig2}, we compare our light curves of PSN~J07285387+3349106 with those published by \citet{tsv15}. The Johnson-Cousins $R$ and $I$-band data
of  \citet{tsv15} are scaled by a constant to match our Sloan $r$ and $i$-band photometry. These zeropoint corrections, $\Delta R = 0.23 \pm 0.03$ mag and $\Delta I = 0.47 \pm 0.03$ mag,
have been computed through synthetic photometry obtained from our spectra. The resulting match between our data and those of Tsvetkov et al. is fairly good.

Early-time unfiltered magnitudes from amateur astronomers, scaled to the Sloan $r$ and the Johnson $V$ systems, allow us to estimate the epochs and the apparent
magnitudes at maximum in these 2 bands. Using a low-order polynomial fit to the light curves, we obtain: $V_{max}$ = 16.63 $\pm$ 0.18 (on JD$_{max}(V)$ = 2457072.0 $\pm$ 0.8)  
and $r_{max}$ = 16.45 $\pm$ 0.15 (on JD$_{max}(r)$ = 2457071.9 $\pm$ 1.1).
The rise time to maximum in the above two bands amounts to more than $\sim$~9~d. 
After peak, we observe a monotonic decline in all bands, with the light curves in the blue bands fading more rapidly than those
in the red bands. The decline rates are moderate in all bands between  0~d and +10~d from maximum: 
15.3 $\pm$ 0.3 mag $100\,\mathrm{d}^{-1}$ in the $u$ band, 9.4 $\pm$ 0.7 mag $100\,\mathrm{d}^{-1}$ in the $V$ band, and
6.7 $\pm$ 1.1 mag $100\,\mathrm{d}^{-1}$ in the $r$ band. From +10~d onwards, the declines become much steeper, 
and are between 28 and 30 mag $100\,\mathrm{d}^{-1}$ in all bands. The average decline rate of PSN~J07285387+3349106 in the $r$ band 
from the light curve peak to the last detection ($\sim$ 22~d after maximum) is 19.7 $\pm$ 1.6 mag $100\,\mathrm{d}^{-1}$, which is
the fastest decline ever observed for a Type Ibn SN in that temporal range.

In Figure  \ref{figabs} (top panel), we show the $B-V$ colour evolution of PSN~J07285387+3349106 for  two different assumptions for the reddening, i.e. E($B-V$)$_{tot}$ = 0.99 mag and E($B-V$)$_{tot}$ = 0.81 mag. 
The colour curves are compared with those of a selected sample of Type Ibn SNe\footnote{The distance modulus and reddening values adopted for the comparison object are the following: $\mu$ = 36.59 mag and E(B-V) = 0.14 mag 
for ASASSN-15ed; $\mu$ = 35.27 mag and E(B-V) = 0.15 mag for SN~1999cq; $\mu$ = 35.52 mag and E(B-V) = 0.11 mag for SN~2000er; $\mu$ = 32.01 mag and E(B-V) = 0.04 mag for SN~2006jc; $\mu$ = 34.27 mag and E(B-V) = 
0.06 mag for SN~2010al.}, 
including the prototype SN~2006jc and a few objects that share spectroscopic and photometric similarities with PSN~J07285387+3349106.
While SN~2006jc is blue and shows very little $B-V$  colour evolution \citep[see discussion in][]{pasto15b}, other objects 
of the sample become redder with time up to about four weeks past maximum. Their colour rises from $B-V \approx$ 0 around maximum to about 1 mag
four weeks later. 

Adopting E($B-V$)$_{tot}$ = 0.99 mag, this would make PSN~J07285387+3349106  slightly bluer than similar objects. 
For this reason, we compute a colour template for Type Ibn SNe through a low-order polynomial fit to the colours of the sample considered in Figure \ref{figabs}, 
from which we remove SN~2006jc because of its different colour evolution.
The best colour match between PSN~J07285387+3349106 and the template is obtained with E($B-V$)$_{tot}$ = 0.81 $\pm$ 0.21 mag.
However, as discussed in Section \ref{reddening}, interacting SNe such as Type Ibn events may easily have intrinsically different colours, so 
the colour match among the objects of this sub-sample cannot be used as a conclusive argument to support a lower reddening scenario for PSN~J07285387+3349106. 
For this reason, in Section \ref{reddening}, we conservatively adopted the weighted average of three independent methods as our best reddening estimate, i.e.   E($B-V$)$_{tot}$ = 0.99 $\pm$ 0.48 mag.

\begin{figure*}
{\includegraphics[width=11.4cm,angle=270]{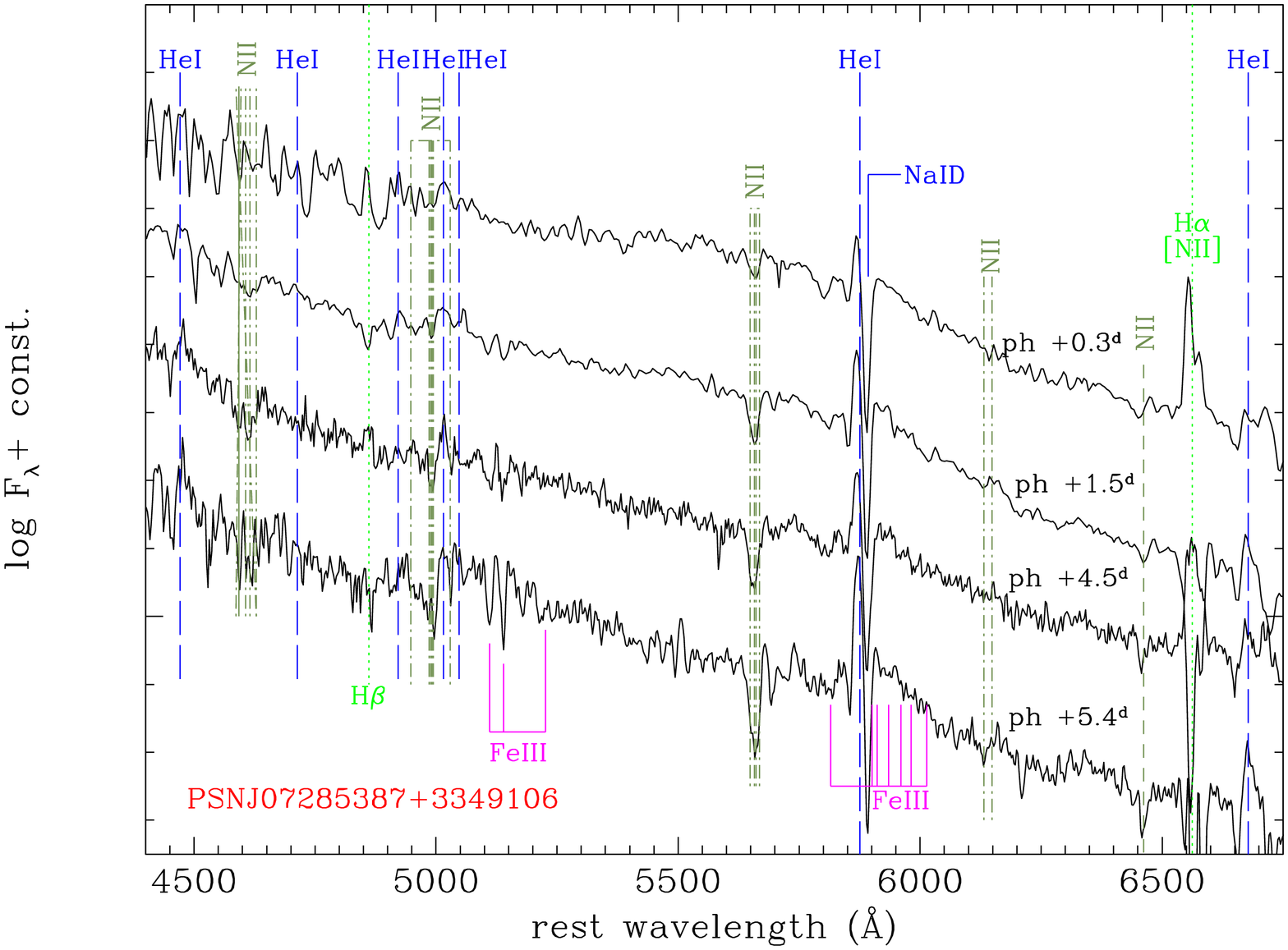}
 \includegraphics[width=11.4cm,angle=270]{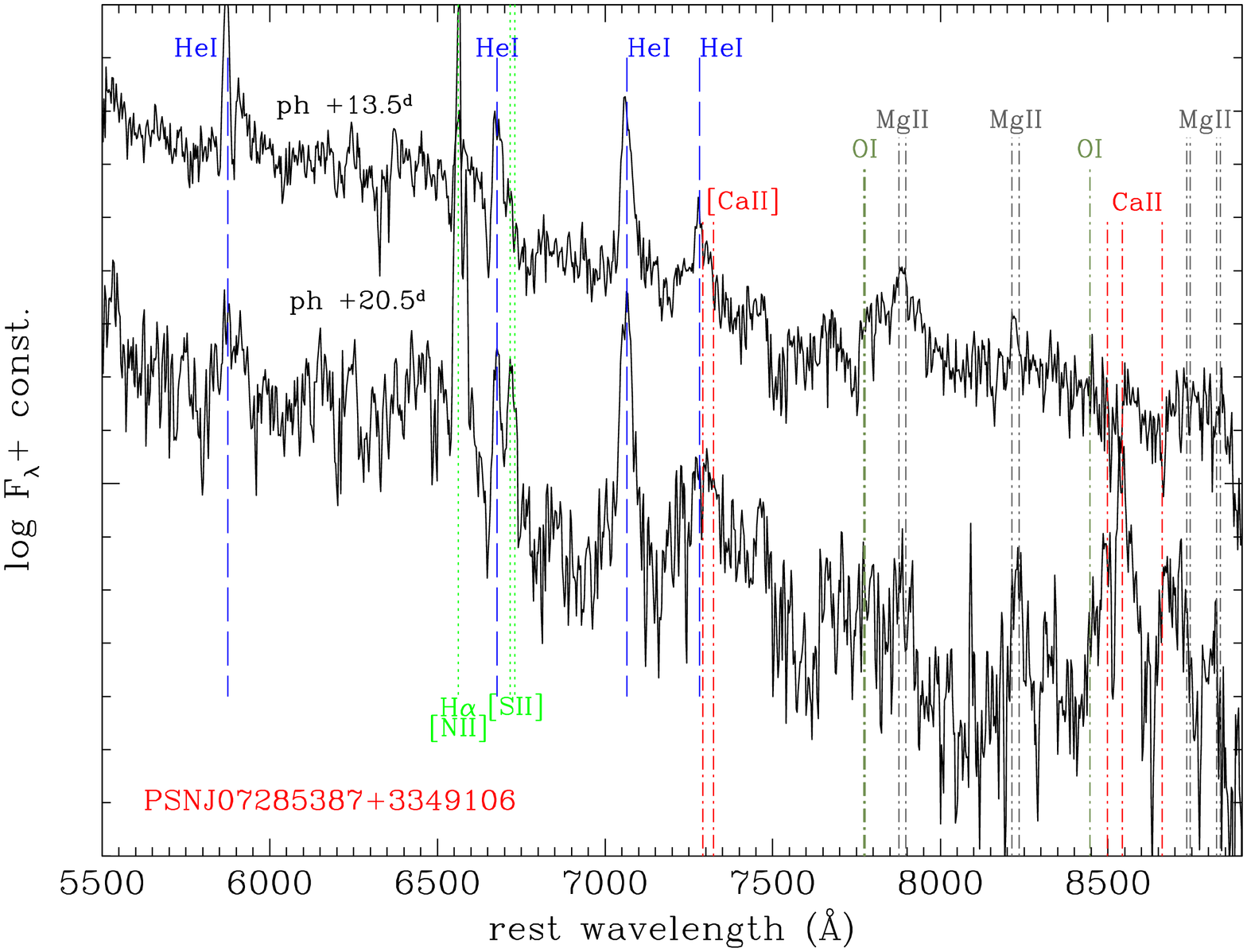}}
 \caption{Top - Line identification in early spectra of PSN~J07285387+3349106.
The Fe~III and N~II lines are marked at the position of the absorption minimum, blueshifted by 950 km s$^{-1}$ from the rest wavelength position. The He~I lines are indicated
at rest wavelength. Bottom - Line identification in late spectra of PSN~J07285387+3349106. All lines are indicated at the rest wavelength position.
The spectra have been corrected for reddening and redshift, using the values adopted in this paper.
 We also mark the H$\alpha$ plus [N~II] blend, H$\beta$, and the [S~II] $\lambda\lambda$6717,6731 doublet, which are due to foreground contamination, with green dotted lines. 
\label{id_early}}
\end{figure*}

Accounting for the distance estimate reported in Section \ref{host}, and the above reddening value, we obtain the following absolute peak
magnitudes in the two bands monitored around the light curve maximum: $M_V$ = $-$20.30 $\pm$ 1.50 and $M_r$ = $-$19.83 $\pm$ 1.19.
A comparison of the absolute $R$-band magnitudes for the same SN sample as before \citep[with the inclusion of SN~1999cq,][]{mat00} is shown in
the bottom panel of Figure  \ref{figabs}. For PSN~J07285387+3349106, we adopted the correction of 0.23 mag 
to convert Sloan r (ABmag) to Johnson-Cousins R (Vegamag) as mentioned in the caption of Figure \ref{fig2}. 
The SN competes with the most luminous objects in the sample.
This comparison shows that there is some similarity in the evolution (in particular, in the decline rates)
among the objects that belong to this restricted sample. However, we remark that the general photometric behaviour of SNe~Ibn can
be strongly heterogeneous. This issue will be discussed in a forthcoming paper \citep{pasto15e}.

\section{Spectroscopic evolution} \label{spec}

The sequence of spectra collected for PSN~J07285387+3349106 is shown in Figure \ref{figspec}.
We  note that there is some residual background contamination in several spectra. A narrow and unresolved H$\alpha$ emission feature due to 
unrelated sources in the host galaxy is prominent in spectra obtained with the lowest resolution grisms or under poor seeing conditions. In other cases,
 a residual absorption is visible at the position of the contaminating H$\alpha$ line due to an over-subtraction of the background.

The early spectra of PSN~J07285387+3349106 show a red, almost featureless continuum, where the most prominent features are due to host galaxy background contamination
(H$\alpha$, [N~II] and [S~II] features) which cannot be properly removed. A narrow, unresolved absorption attributed to the Na~ID feature
is also clearly detected in our early spectra. This is produced by Na~I atoms lying along the line of sight inside NGC~2388. 
However, there is no clear evidence that this absorption is related the SN circumstellar environment. 
The strength of the interstellar Na~ID absorption decreases with time, because a prominent narrow He~I $\lambda$5876 emission feature emerges in its blue wing.
Narrow emission lines of He~I $\lambda$6678 (blended, in some early spectra, with the host galaxy [S~II] doublet) and $\lambda$7065 are also detected with P-Cygni profile.

We attempt to identify a number of weak lines visible in the early-phase spectra (see Figure \ref{id_early}, top). In particular, we suggest the unusual 
detection of Fe~III and N~II features, with absorption components blueshifted by about 950 km s$^{-1}$. In particular, Fe~III $\lambda$5127 and  Fe~III $\lambda$5156 are strong. The identification of Fe~III lines at 5800-6000~\AA~is ambiguous given the low S/N level.
N~II blends likely produce some absorption features at about 4620~\AA, 5650~\AA~and 6130~\AA. In addition, we identify N~II $\lambda$6482.
Although these lines have not been identified before in early spectra of SNe Ibn,
they have occasionally been detected in massive, hot stars \citep[e.g.,][]{gva10}.

In our late spectra, He~I $\lambda$7065 appears to be slightly broader, with a full width at half maximum (FWHM) velocity increasing to about 1300 km s$^{-1}$.
In addition, whilst the continuum becomes even redder, other lines appear at the long wavelengths, likely due to O~I ($\lambda$7774--7777 and $\lambda$8446), Mg~II ($\lambda$7877--7896, $\lambda$8214--8235, $\lambda$8735--8745, $\lambda$8824--8835) and Ca~II. In particular, the NIR Ca~II triplet 
becomes the strongest SN feature visible in our latest spectrum, while [Ca~II] $\lambda$$\lambda$7291,7324 is barely detected, blended with He~I $\lambda$7281. 
Unfortunately, the large reddening, the heavy background contamination and the relatively
modest S/N of our spectra do not allow to securely identify other spectral lines.

\begin{figure}
\includegraphics[width=8.5cm,angle=0]{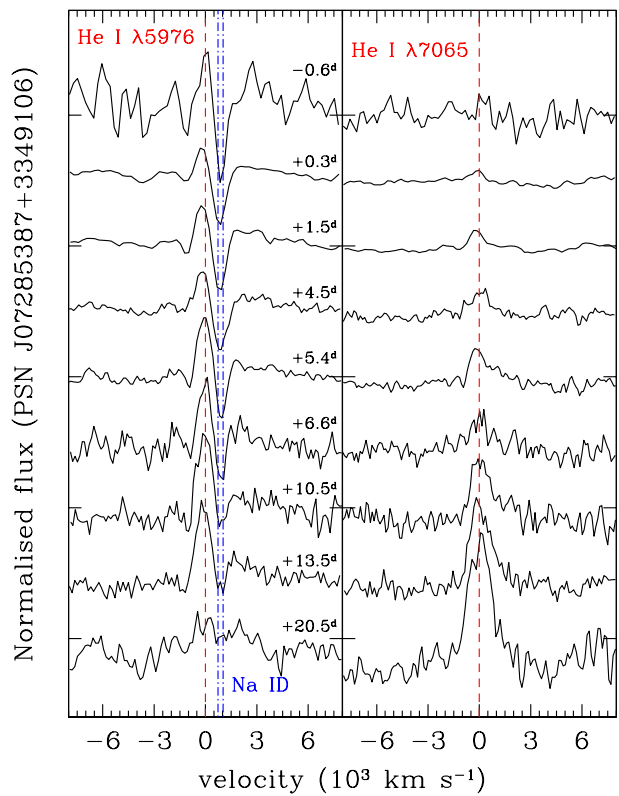} 
 \caption{Evolution of the He~I $\lambda$5876 and Na~ID blend (left) and  the He~I $\lambda$7065 line (right) in our spectra of PSN~J07285387+3349106.  The vertical dashed lines mark the rest velocity of the He~I lines, while the two dot-dashed lines in the left panel mark the position of the two
Na~ID absorption lines. The velocity scale on the abscissa is in units of 10$^3$ km s$^{-1}$.
\label{fig_He}}
\end{figure}

\begin{figure}
\includegraphics[width=8.8cm,angle=0]{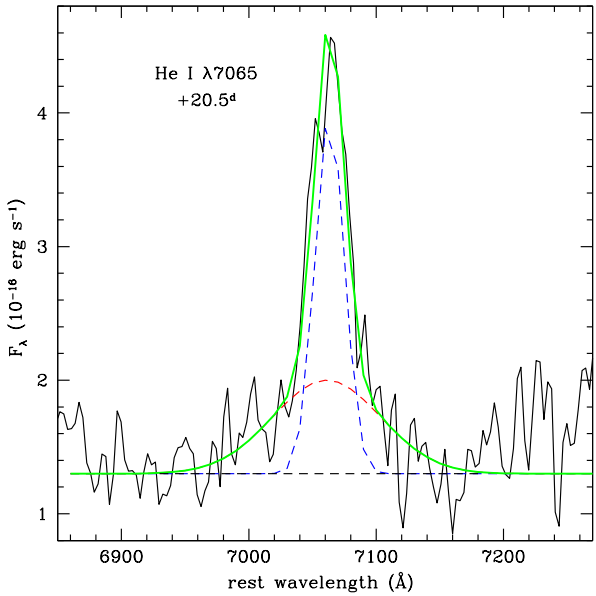} 
 \caption{Two-component Gaussian fit to the He~I $\lambda$7065 line in the latest NOT + ALFOSC spectrum (phase +20.5 d).
\label{fig7}}
\end{figure}

The evolution of the He~I $\lambda$5876 plus Na~ID blend (left panel) and the He~I $\lambda$7065 line (right panel) is presented in Figure \ref{fig_He}.
We note the increasing strength of the He~I $\lambda$5876 line relative to the Na~ID foreground absorption, which almost completely disappears in our late spectra.
Similarly, the He~I $\lambda$7065 line increases in strength, becoming one of the most prominent SN features in our late spectra.
The velocity of the slow-moving
gas, as measured from the FWHM of the narrow He~I $\lambda$7065 component or from the position of its weak narrow P-Cygni absorption, 
is about 900-1000 km s$^{-1}$. 
A broader component is also marginally detected
in the  He~I $\lambda$7065 feature, with an approximate velocity of about 3100-3400 km s$^{-1}$. This component is best seen in our late time spectra
 (see e.g. Figure \ref{fig7}).

\section{Discussion} \label{disc}

The analysis of the spectroscopic data of PSN~J07285387+3349106 suggests that the very red apparent colour detected at all epochs is a consequence of 
a relatively large reddening in the direction to the SN. Once corrected for the intrinsic extinction, PSN~J07285387+3349106 is a Type Ibn SN with
similar properties as already observed in a number of objects of this type. 

In particular, the spectrum at early phases is dominated by narrow P-Cygni lines of He~I. We propose that these features, 
with a FWHM velocity of about 900-1000 km s$^{-1}$, are the signature of a 
circumstellar wind. The velocity of this CSM, along with its chemical composition, may provide
important insights into the nature of the progenitor star. This wind is likely H-free, because the unresolved Balmer lines observed
in the spectra are attributed to host galaxy contamination. 

Specific for this object and unprecedented  is the detection of N~II lines in the early spectra. This 
prompts for a WN-type Wolf-Rayet progenitor. More specifically, the lack of circumstellar Balmer lines suggests a WNE subtype classification \citep{ham95}. This subclass of Wolf-Rayet stars
has a typical terminal wind velocity of about 1600 km s$^{-1}$ \citep{ham06}, consistent with the velocities measured for the bulk of the circumstellar material of PSN~J07285387+3349106.

PSN~J07285387+3349106 is also a very luminous event (likely brighter than $-$20 mag, though this estimate is affected by a big uncertainty in the light-of-sight extinction), showing an extremely fast
post-peak decline, without a visible flattening onto the $^{56}$Co tail. All of this suggests that the luminous peak of the light curve is powered
by interaction with CSM rather than a large amount of synthesized $^{56}$Ni.

A number of authors \citep[e.g.,][]{ofe10,che11,svi12} suggest that a fraction of interacting SNe have early light curves powered by shock breakout in a dense wind.
In fact, while the shock breakout emerging from the stellar surface has a very short duration (of the order of minutes) and releases UV and X-ray photons, 
the situation changes when the SN explodes within a dense circumstellar cocoon. In that case, the shock may propagate in the CSM without immediately releasing photons to the observer. 
By the time the forward shock breaks out, it has accumulated a lot of energy which is released over much longer timescales than in 
the windless case.
This mechanism may explain the luminous and long-lasting light curve peaks of a number of interacting SNe embedded in a dense CSM, as suggested by \citet{ofe14a}
\citep[but see][]{mor14}. The shock breakout model in a dense circumstellar wind has the advantage of explaining the enormous peak luminosity of PSN~J07285387+3349106,
along with its relatively rapid photometric evolution with respect to the very slow evolutionary time scales of more extreme interacting SNe \citep[e.g.][]{are99}.

A potential problem of this interpretation is the requirement of high CSM densities. In order to produce such a dense CSM, the precursor star should have high mass-loss rate, from a few 10$^{-3}$ M$_\odot$ yr$^{-1}$ to
10$^{-1}$ M$_\odot$ yr$^{-1}$ \citep{kie12,tad13,mor14b}. These substantial mass-loss rates are expected in luminous blue variable (LBV) eruptions, and not in steady winds of the putative Wolf-Rayet progenitors of SNe Ibn \citep{ham00,nug00}.
Nonetheless, at least in the case of SN~2006jc, a residual LBV-like behaviour has been observed through the detection of a pre-SN burst in 2004 \citep{nak06,pasto07}.
A rather complex CSM density profile and a variable mass-loss rate, ranging from 10$^{-1}$ M$_\odot$ yr$^{-1}$ during the outburst to many orders of magnitude less
before and after the 2004 episode, were proposed by \citet{tom08}. We emphasise that the shock breakout scenario in an optically thick CSM has also been suggested for the Type Ibn SN
iPTF13beo \citep{gor14}. The shock breakout interpretation for this SN requires an enormous mass-loss (about 2.4  M$_\odot$ yr$^{-1}$) over a very short period. 
In other words, though unobserved, an LBV-like violent eruption is necessary also to explain the properties of iPTF13beo. 
The ultimate verification of this scenario is the detection of pre-SN eruptions. So far, SN~2006jc is the only SN Ibn for which a pre-SN eruption has been registered. However, 
pre-SN observations of the host galaxies are not available for most Type Ibn SNe, or these SNe are too distant to detect $\sim -$14 mag pre-SN transient events. 

It is evident that routine observations of so-called ``SN impostors'' \citep{van00,mau06,pasto10,pasto13,tar15a,kan15} are essential to monitor extremely rare very massive stars that 
are approaching their final death as luminous SNe. This strategy is complemented by panoramic surveys such as the {\it intermediate Palomar Transient Factory}, 
{\it La Silla-QUEST}, {\it Pan-STARRS}, the {\it Catalina Real-Time Transient Survey}, and the {\it SkyMapper Southern Sky Survey} which provide extensive 
databases of future SN explosion sites. These images will allow us to recover a posteriori the variability history of the progenitors of new interacting SNe, as done for some recent SNe IIn
\citep{fra13,ofe14b,tar15b}.

\section*{Acknowledgements}

We are grateful to Massimo Conti, Giacomo Guerrini, Paolo Rosi, Luz Marina Tinjaca Ramirez for their help with the observations. 
We are also grateful to Manfred Mrotzek ({\it http://www.astro-photos.net/index$\_$en.htm}) and Paolo Campaner ({\it http://paolocampaner.blogspot.it}) for providing
their observations of PSN~J07285387+3349106.
Part of the observations in Asiago were done during the Opticon/Neon school
({\it www.iap.fr/neon/}), by the PhD students Demetra de Cicco (Napoli), Marco Lam
(Edimburgh), Ignacio Ordovas (Santander) and Paulina Sowicka (Warsaw), under the
supervision of Nancy Elias-Rosa, and Michel Dennefeld. Opticon is funded by the
European Commission's FP7 Capacities programme (Grant number 312430).

AP, SB, NER, AH, LT, GT, and MT are partially supported by the PRIN-INAF 2014 with the project ''Transient Universe: unveiling 
new types of stellar explosions with PESSTO''. 
NER acknowledges the support from the European Union Seventh Framework Programme (FP7/2007-2013) under grant agreement n. 267251 ''Astronomy Fellowships in Italy'' (AstroFIt). 
AMG acknowledges financial support by the Spanish Ministerio de Econom\'ia y Competitividad (MINECO) grant ESP2013-41268-R.
ST and UMN acknowledge support by TRR33 ``The Dark Universe'' of the German Research Foundation (DFG).
 EEOI is partially supported by the Brazilian agency CAPES (grant number 9229-13-2).
We gratefully acknowledge the support from the  Knut and Alice Wallenberg Foundation. The Oskar Klein Centre is funded by the Swedish Research Council. 

This paper is based on observations made with the Italian Telescopio Nazionale Galileo
(TNG) operated on the island of La Palma by the Fundaci\'on Galileo Galilei of
the INAF (Istituto Nazionale di Astrofisica) and the Copernico Telescope of INAF-Asiago Observatory and the Rapid Eye Mount (REM) INAF telescope
hosted at La-Silla (European Southern Observatory, Chile). 
This work is also based on observations obtained with the Nordic Optical Telescope, which  is operated by the 
Nordic Optical Telescope Scientific Association at the Observatorio del Roque de los Muchachos, La Palma, Spain, of the Instituto de Astrofisica de Canarias. 
This paper is also based on observations obtained with the Joan Or\'o
Telescope (TJO) of the Montsec Astronomical Observatory (OAdM) which is
owned by the Catalan Government and operated by the Institute for Space
Studies of Catalonia (IEEC).

This research has made use of the NASA/IPAC Extragalactic Database (NED) which is operated by the Jet Propulsion Laboratory, California Institute of Technology, 
under contract with the NASA. We acknowledge the usage of the HyperLeda database (http://leda.univ-lyon1.fr).

\bsp

\label{lastpage}

\end{document}